\documentclass[twocolumn,showpacs,aps,prl,superscriptaddress]{revtex4}

\usepackage{graphicx}
\usepackage{dcolumn}
\usepackage{amsmath}
\usepackage{epsfig}
\RequirePackage{xspace}

\usepackage{relsize}
\def\babar{\mbox{\slshape B\kern-0.1em{\smaller A}\kern-0.1em
    B\kern-0.1em{\smaller A\kern-0.2em R}}}

\def\electron   {\ensuremath{e}\xspace}
\def\en         {\ensuremath{e^-}\xspace}   % electron negative (\em is taken)
\def\ep         {\ensuremath{e^+}\xspace}
 
\def\epem       {\ensuremath{e^+e^-}\xspace}

\def\mup        {\ensuremath{\mu^+}\xspace}
\def\mun        {\ensuremath{\mu^-}\xspace} % muon negative (\mum is taken)

\def\ellm       {\ensuremath{\ell^-}\xspace}
\def\ellp       {\ensuremath{\ell^+}\xspace}

\def\q     {\ensuremath{q}\xspace}

\def\qqbar {\ensuremath{q\overline q}\xspace}
\def\u     {\ensuremath{u}\xspace}

\def\d     {\ensuremath{d}\xspace}

\def\s     {\ensuremath{s}\xspace}

\def\c     {\ensuremath{c}\xspace}

\def\piz   {\ensuremath{\pi^0}\xspace}

\def\Kbar  {\kern 0.2em\overline{\kern -0.2em K}{}\xspace}

\def\Kz    {\ensuremath{K^0}\xspace}
\def\Kzb   {\ensuremath{\Kbar^0}\xspace}
\def\KzKzb {\ensuremath{\Kz \kern -0.16em \Kzb}\xspace}
\def\Kp    {\ensuremath{K^+}\xspace}
\def\Km    {\ensuremath{K^-}\xspace}

\def\KpKm  {\ensuremath{\Kp \kern -0.16em \Km}\xspace}
\def\KS    {\ensuremath{K^0_{\scriptscriptstyle S}}\xspace}

\def\Dbar    {\kern 0.2em\overline{\kern -0.2em D}{}\xspace}

\def\Dz      {\ensuremath{D^0}\xspace}
\def\Dzb     {\ensuremath{\Dbar^0}\xspace}
\def\DzDzb   {\ensuremath{\Dz {\kern -0.16em \Dzb}}\xspace}
\def\Dp      {\ensuremath{D^+}\xspace}
\def\Dm      {\ensuremath{D^-}\xspace}

\def\DpDm    {\ensuremath{\Dp {\kern -0.16em \Dm}}\xspace}

\def\B       {\ensuremath{B}\xspace}
\def\Bbar    {\kern 0.18em\overline{\kern -0.18em B}{}\xspace}

\def\BB      {\ensuremath{B\Bbar}\xspace} 
\def\Bz      {\ensuremath{B^0}\xspace}
\def\Bzb     {\ensuremath{\Bbar^0}\xspace}
\def\BzBzb   {\ensuremath{\Bz {\kern -0.16em \Bzb}}\xspace}
\def\Bu      {\ensuremath{B^+}\xspace}
\def\Bub     {\ensuremath{B^-}\xspace}

\def\BpBm    {\ensuremath{\Bu {\kern -0.16em \Bub}}\xspace}

\def\BorBbar    {\kern 0.18em\optbar{\kern -0.18em B}{}\xspace}
\def\DorDbar    {\kern 0.18em\optbar{\kern -0.18em D}{}\xspace}
\def\KorKbar    {\kern 0.18em\optbar{\kern -0.18em K}{}\xspace}

\def\jpsi     {\ensuremath{{J\mskip -3mu/\mskip -2mu\psi\mskip 2mu}}\xspace}

\mathchardef\Upsilon="7107
\def\Y#1S{\ensuremath{\Upsilon{(#1S)}}\xspace}% no space before {...}!

\def\FourS {\Y4S}

\mathchardef\Deltares="7101
\mathchardef\Xi="7104
\mathchardef\Lambda="7103
\mathchardef\Sigma="7106
\mathchardef\Omega="710A

\def\Deltabar{\kern 0.25em\overline{\kern -0.25em \Deltares}{}\xspace}
\def\Lbar{\kern 0.2em\overline{\kern -0.2em\Lambda\kern 0.05em}\kern-0.05em{}\xspace}
\def\Sigbar{\kern 0.2em\overline{\kern -0.2em \Sigma}{}\xspace}
\def\Xibar{\kern 0.2em\overline{\kern -0.2em \Xi}{}\xspace}
\def\Obar{\kern 0.2em\overline{\kern -0.2em \Omega}{}\xspace}
\def\Nbar{\kern 0.2em\overline{\kern -0.2em N}{}\xspace}
\def\Xb{\kern 0.2em\overline{\kern -0.2em X}{}\xspace}

\def\mes        {\mbox{$m_{\rm ES}$}\xspace}

\def\DeltaE     {\mbox{$\Delta E$}\xspace}

\newcommand{\tev}{\ensuremath{\mathrm{\,Te\kern -0.1em V}}\xspace}
\newcommand{\gev}{\ensuremath{\mathrm{\,Ge\kern -0.1em V}}\xspace}
\newcommand{\mev}{\ensuremath{\mathrm{\,Me\kern -0.1em V}}\xspace}
\newcommand{\kev}{\ensuremath{\mathrm{\,ke\kern -0.1em V}}\xspace}
\newcommand{\ev}{\ensuremath{\mathrm{\,e\kern -0.1em V}}\xspace}
\newcommand{\gevc}{\ensuremath{{\mathrm{\,Ge\kern -0.1em V\!/}c}}\xspace}
\newcommand{\mevc}{\ensuremath{{\mathrm{\,Me\kern -0.1em V\!/}c}}\xspace}
\newcommand{\gevcc}{\ensuremath{{\mathrm{\,Ge\kern -0.1em V\!/}c^2}}\xspace}
\newcommand{\mevcc}{\ensuremath{{\mathrm{\,Me\kern -0.1em V\!/}c^2}}\xspace}
\def\mus  {\ensuremath{\rm \,\mus}\xspace}

\def\ps   {\ensuremath{\rm \,ps}\xspace}

\def\mus        {\ensuremath{\,\mu{\rm s}}\xspace}    %% microsecond
\def\ps         {\ensuremath{{\rm \,ps}}\xspace}  %% picosecond

\def\to                 {\ensuremath{\rightarrow}\xspace}

\newcommand{\stat}{\ensuremath{\mathrm{(stat)}}\xspace}
\newcommand{\syst}{\ensuremath{\mathrm{(syst)}}\xspace}

\def\pep2{PEP-II}

\def\gsim{{~\raise.15em\hbox{$>$}\kern-.85em
          \lower.35em\hbox{$\sim$}~}\xspace}
\def\lsim{{~\raise.15em\hbox{$<$}\kern-.85em
          \lower.35em\hbox{$\sim$}~}\xspace}

\def\CP                {\ensuremath{C\!P}\xspace}
\def\C       {\ensuremath{C}\xspace}

\def\deltat{\ensuremath{{\rm \Delta}t}\xspace}
\def\deltamd{\ensuremath{{\rm \Delta}m_d}\xspace}

\xspace

\newcommand{\jprlBase}       {Phys.\ Rev.\ Lett.\xspace}
\newcommand{\jprBase}        {Phys.\ Rev.\xspace}
\newcommand{\jplBase}        {Phys.\ Lett.\xspace}
\newcommand{\nimBaseA}       {Nucl.\ Instr.\ Methods Phys.\ Res., Sect.\ A\xspace}

\newcommand{\npBase}         {Nucl.\ Phys.\xspace}

\newcommand{\nima}      [1]  {\nimBaseA~{\bf #1}}

\newcommand{\np}        [1]  {\npBase\ {\bf #1}}
\newcommand{\plb}       [1]  {\jplBase\ B~{\bf #1}}

\newcommand{\jprl}      [1]  {\jprlBase\ {\bf #1}}
\newcommand{\pr}        [1]  {\jprBase\ {\bf #1}}
\newcommand{\jprd}      [1]  {\jprBase\ D~{\bf #1}}
\def\jetset74   {\mbox{\tt Jetset \hspace{-0.5em}7.\hspace{-0.2em}4}\xspace}
\def\sintwob    {\ensuremath{\sin 2 \beta}}

\def\fish    {\ensuremath{\cal F}}
\def\btodstarrho {\ensuremath{\B \to {D}^{\star}\:\rho}}

\def\btojpsipi {\ensuremath{\B^0 \to \jpsi\piz}}
\def\btojpsiks {\ensuremath{\B^0 \to \jpsi\KS}}
\def\kstopizpiz {\ensuremath{\KS \to \piz\piz}}

\def\jpsitoll {\ensuremath{\jpsi\rightarrow\ellp\ellm}}
\def\jpsitoee {\ensuremath{\jpsi\rightarrow\ep\en}}
\def\jpsitomm {\ensuremath{\jpsi\rightarrow\mup\mun}}

\def\C {\ensuremath{C}}
\def\S {\ensuremath{S}}

\def\dt   {\ensuremath{\Delta t}}

\def\dz   {\ensuremath{\Delta z}}

\def\d    {\ensuremath{d}}
\def\u    {\ensuremath{u}}
\def\s    {\ensuremath{s}}

\def\bb   {\ensuremath{B\Bbar}}

\def\cerenkov{$\check{\rm C}{\rm erenkov}$ }

\newcommand{\BABARPubYear}    {05}
\newcommand{\BABARPubNumber}  {54}

\newcommand{\SLACPubNumber} {11731}

\def\figurebox#1#2#3{%
    \def\arg{#3}%
    \ifx\arg\empty
    {\hfill\vbox{\hsize#2\hrule\hbox to #2{\vrule\hfill\vbox to #1{\hsize#2\vfill}\vrule}\hrule}\hfill}%
    \else
    {\hfill\epsfbox{#3}\hfill}%
    \fi}

\begin{document}

\preprint{\babar-PUB-\BABARPubYear/\BABARPubNumber} 
\preprint{SLAC-PUB-\SLACPubNumber} 

\begin{flushleft}
\babar-PUB-\BABARPubYear/\BABARPubNumber\\
SLAC-PUB-\SLACPubNumber\\
%hep-ex/\LANLNumber\\[10mm]\\
%\rm \babar$\;$ Analysis Document \# 1322\\
%Version 8
\end{flushleft}

\title{
{\large \bf
Measurements of the Branching Fraction and Time-Dependent CP Asymmetries of {\boldmath $\btojpsipi$} decays} 
}

% Use the December 2005 author list
%\input pubboard/authors_dec2005
%% author list as of 02-Dec-2005 (621 authors)
%
\author{B.~Aubert}
\author{R.~Barate}
\author{D.~Boutigny}
\author{F.~Couderc}
\author{Y.~Karyotakis}
\author{J.~P.~Lees}
\author{V.~Poireau}
\author{V.~Tisserand}
\author{A.~Zghiche}
\affiliation{Laboratoire de Physique des Particules, F-74941 Annecy-le-Vieux, France }
\author{E.~Grauges}
\affiliation{IFAE, Universitat Autonoma de Barcelona, E-08193 Bellaterra, Barcelona, Spain }
% ...
\collaboration{The \babar\ Collaboration}
\noaffiliation

\date{\today}% It is always \today, today, but you may specify any date with \date.

%
% Abstract
%
\begin{abstract}
We present measurements of the branching fraction and time-dependent \CP\ asymmetries 
in\newline $\btojpsipi$ decays based on (231.8 $\pm$ 2.6) $\times$ 10$^{6}$ 
$\FourS\rightarrow\bb$ decays collected with the \babar\ detector at the SLAC PEP-II 
asymmetric-energy \B\ factory. We obtain a branching fraction 
$\cal{B}$($\btojpsipi$) = (1.94 $\pm$ 0.22 \stat $\pm$ 0.17 \syst)$\times$ 10$^{-5}$. 
We also measure the \CP\ asymmetry parameters \C\ = $-$0.21 $\pm$ 0.26 \stat $\pm$ 0.06 \syst 
and \S\ = $-$0.68 $\pm$ 0.30 \stat $\pm$ 0.04 \syst.
\end{abstract}

% PACS, the Physics and Astronomy Classification Scheme.

\maketitle

Charge conjugation-parity (\CP) violation in the $B$ meson system has been
established by the \babar~\cite{babar-stwob-prl}
and Belle~\cite{belle-stwob-prl} collaborations.
The Standard Model (SM) of electroweak interactions describes \CP\ violation
as a consequence of a complex phase in the
three-generation Cabibbo-Kobayashi-Maskawa (CKM) quark-mixing
matrix~\cite{ref:CKM}. Measurements of \CP\ asymmetries in
the proper-time distribution of neutral $B$ decays to
\CP\ eigenstates containing a charmonium and $K^{0}$ meson provide
a precise measurement of $\sintwob$~\cite{BCP}, where
$\beta$ is $\arg \left[\, -V_{\rm cd}^{}V_{\rm cb}^* / V_{\rm td}^{}V_{\rm tb}^*\, \right]$ and
the $V_{ij}$ are CKM matrix elements. 

\par
\begin{figure}[h!]
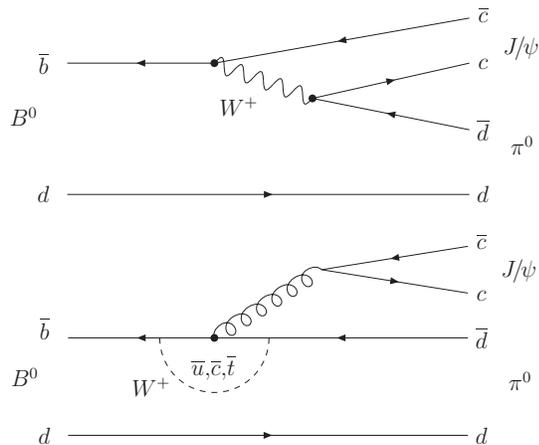

\begin{center}
\includegraphics*[height=3.0cm]{jpsipizTree.ps_edit}
\includegraphics*[height=3.5cm]{jpsipizPeng.ps_edit}
\caption{Feynman diagrams of the color suppressed tree (top) and gluonic penguin (bottom) amplitudes contributing to the \btojpsipi\ decay.}
\label{fig:feynman}
\end{center}
\end{figure}

 The decay $\btojpsipi$ is a \CP-even Cabibbo-suppressed 
${b \rightarrow c\mskip 2mu \overline c \mskip 2mu d}$ transition
whose tree amplitude has the same weak phase as the
${b \rightarrow c\mskip 2mu \overline c \mskip 2mu s}$ modes
{\emph{e.g.}} the \CP-odd decay $\btojpsiks$. The ${b \rightarrow c\mskip 2mu \overline c \mskip 2mu d}$
penguin amplitude has a different weak phase than the tree amplitude.
The tree and penguin amplitudes expected to dominate this decay are shown in Figure~\ref{fig:feynman}. 
%---------------------------
% Figure : Feynman diagrams
%---------------------------
%------------------------------
\par If there is a significant penguin amplitude in $\btojpsipi$, then one will
measure values of the \CP asymmetry coefficients \S\ and \C\ that are different
from $-\sintwob$ and 0, respectively~\cite{grossman}.
%of the \CP asymmetry coefficients \S\ and \C; \S\ that differs 
%from $-\sintwob$, and \C\ that differs from zero~\cite{grossman}. 
The coefficient \S\ denoting the interference between mixing and decay, and 
the direct \CP asymmetry coefficient $\C$ are defined as:
\begin{eqnarray}
S \equiv \frac{{2\mathop{\cal I\mkern -2.0mu\mit m}}
  \lambda}{1+|\lambda|^2}
\mskip 50mu {\rm and} \mskip 50mu
C \equiv \frac{1  - |\lambda|^2 }{1+|\lambda|^2},
\label{equation-sandc}
\end{eqnarray}
where $\lambda$ is a complex parameter that depends on both the \Bz-\Bzb oscillation 
amplitude and the amplitudes describing \Bz and \Bzb decays to the $\jpsi\piz$ final state.
An additional motivation for measuring \S\ and \C\ from $\btojpsipi$ is that 
they can provide a model-independent constraint on the penguin dilution 
within $\btojpsiks$~\cite{ciuchini}.

\par In this publication, we present an update of previous \babar\ 
branching fraction and time-dependent \CP\-violating asymmetry measurements
of the decay $\btojpsipi$~\cite{ref:babarbf,ref:babarcp}, which had 
been performed using 20.7 fb$^{-1}$ and 81.1 fb$^{-1}$ of integrated 
luminosity, respectively. Belle has also studied this mode and has published a branching
fraction and later a time-dependent \CP\-violating asymmetry result using
29.4 fb$^{-1}$ and 140.0 fb$^{-1}$ of integrated luminosity, respectively~\cite{ref:bellebf,ref:bellecp}. 

The data used in this analysis were collected with the \babar\ detector at the \pep2\ asymmetric $\epem$ 
storage ring. This represents a total integrated luminosity 
of 210.6 fb$^{-1}$ collected on or just below the $\FourS$ resonance (on-peak),
 corresponding to a sample of 231.8 $\pm$ 2.6 million $\BB$ pairs. An additional 21.6 fb$^{-1}$ of data, collected 
at approximately 40 MeV below the $\FourS$ resonance, is used to study
background from \epem\to\qqbar ($\q = \u,\d,\s,\c$) continuum events.

The \babar\ detector is described in detail elsewhere~\cite{ref:babar}. 
Surrounding the interaction point is a 5 layer double-sided silicon vertex tracker (SVT) which
%provides precise reconstruction of track angles and \B\ decay vertices. 
measures the impact parameters of charged particle tracks in both the plane transverse
to, and along the beam direction.
A 40 layer drift chamber (DCH) surrounds the SVT and provides measurements
of the transverse momenta for charged particles. Both the SVT and the DCH
operate in the magnetic field of a 1.5 T solenoid. Charged hadron 
% operate in a 1.5 T solenoidal magnetic field. Charged hadron 
identification is achieved through measurements of particle energy loss 
($dE/dx$) in the tracking system and the \cerenkov angle obtained 
from a detector of internally reflected \cerenkov light (DIRC).  This is surrounded by a 
segmented CsI(Tl) electromagnetic calorimeter (EMC) which is used to provide photon 
detection and electron identification, and is used to reconstruct neutral 
hadrons.  Finally, the instrumented flux return (IFR) of the magnet allows 
discrimination of muons from pions.

We reconstruct $\btojpsipi$ decays in $\bb$ candidate events from combinations
 of $\jpsitoll$ ($\ell$ = $\electron$, $\mu$) and $\piz\rightarrow\gamma\gamma$ candidates.
A detailed description of the charged particle reconstruction and identification
can be found elsewhere~\cite{ref:babarbf}. For the $\jpsitoee$ ($\jpsitomm$) channel, 
the invariant mass of the lepton pair is required to be between $3.06$ and $3.12 \gevcc$ 
($3.07$ and $3.13 \gevcc$). Each lepton candidate must also be consistent with the 
electron (muon) hypothesis. We form $\piz\to\gamma\gamma$ candidates from clusters in the EMC with
an invariant mass, $m_{\gamma\gamma}$ satisfying $100 < m_{\gamma\gamma} < 160$ {\mevcc}.
These clusters are required to be isolated from any charged tracks, carry
a minimum energy of 30{\mev}, and have a lateral energy distribution 
 consistent with that of a photon. Each $\piz$ candidate is required to
have a minimum energy of 200{\mev} and is constrained
to the nominal mass~\cite{ref:pdg2004}. Finally the $\btojpsipi$ candidates ($B_{rec}$) are constrained 
to originate from the \epem{} interaction point using a geometric fit. 

We use two kinematic variables, $\mes$ and \DeltaE, in order to isolate the signal:
$\mes=\sqrt{(E_{\rm beam}^{\rm *})^2-(p_B^{\rm *})^2}$ is the beam-energy substituted mass and
\DeltaE = $E_B^{\rm *} - E_{\rm beam}^{\rm *}$ is the difference between the \B-candidate energy 
and the beam energy. $E_{\rm beam}^{\rm *}$ and $p_B^{\rm *}$ ($E_B^{\rm *} $) are the beam energy and 
\B-candidate momentum (energy) in the center-of-mass (CM) frame. We require 
$\mes > 5.2 \gevcc$ and $|\DeltaE| < 0.3 \gev$.

\par A significant source of background is due to \epem\to\qqbar ($\q = \u,\d,\s,\c$) 
continuum events.  We combine several kinematic and topological variables 
into a Fisher discriminant (\fish)~\cite{ref:fisher} to provide additional separation between signal
 and continuum. The three variables $L_0$, $L_2$ and $\cos$($\theta_{H}$) are inputs to \fish.
$L_0$ and $L_2$ are the zeroth- and second-order Legendre polynomial moments;
%$L_0 = \sum_i |{\bf p}^{\rm *}_i|$ and $L_2 = \sum_i |{\bf p}^{\rm *}_i| \hspace{0.5mm} \frac{3 \cos^2\theta_i - 1}{2}$,
$L_0 = \sum_i |{\bf p}^{\rm *}_i|$ and $L_2 = \sum_i |{\bf p}^{\rm *}_i|/2 \hspace{0.5mm} {(3 \cos^2\theta_i - 1)}$,
where ${\bf p}^{\rm *}_i$ are the CM momenta of the tracks and neutral
calorimeter clusters that are not associated with the signal candidate. The
$\theta_i$ are the angles between ${\bf p}^{\rm *}_i$ and the thrust axis of
the signal candidate and $\theta_{H}$ is the angle between the positively charged lepton 
and the $\B$ candidate in the $\jpsi$ rest frame.
%$\theta_{H}$ is the angle between one of the leptons 
%and the $\B$ candidate in the $\jpsi$ rest frame.

\par We use multivariate algorithms to identify signatures of \B\ decays that determine (tag)
the flavor of the decay of the other \B in the event ($\B_{tag}$) to be either a \Bz or \Bzb. The flavor 
tagging algorithm used is described in more detail elsewhere~\cite{ref:babar2004}.
In brief, we define seven mutually exclusive tagging categories.  These are 
(in order of decreasing signal purity) Lepton, 
KaonI, KaonII, Kaon-Pion, Pion, Other, and No-Tag.
The total effective tagging efficiency of this algorithm is ($30.5 \pm 0.4$)\%.

%\par We use multivariate algorithms to identify signatures of \B\ decays that determine (tag)
%the flavor of the decay of the other \B in the event ($\B_{tag}$) to be either a \Bz or \Bzb. The flavor 
%tagging algorithm used has a total effective efficiency of $30.5 \pm 0.4$\%  and 
%is described in more detail elsewhere~\cite{ref:babar2004}.

\par The decay rate $f_+$ ($f_-$) of neutral decays to a \CP eigenstate, 
when $B_{tag}$ is a  \Bz (\Bzb), is:
\begin{equation}
f_{\pm}(\dt) = \frac{e^{-\left|\dt\right|/\tau_{\Bz}}}{4\tau_{\Bz}} [1
\pm S\sin(\deltamd\deltat) \mp \C\cos(\deltamd\dt)],  
\label{equation-ff}
\end{equation}
where \dt\ is the difference between the proper decay times of 
the $B_{rec}$ and $B_{tag}$ mesons, $\tau_{\Bz}$ = 1.536 $\pm$ 0.014 ps is the \Bz\ lifetime
and \deltamd\ = 0.502 $\pm$ 0.007 ps$^{-1}$ is the \Bz-\Bzb\ oscillation frequency~\cite{ref:pdg2004}. 
The decay width difference between the \Bz\ mass eigenstates is assumed to be zero. 

\par The time interval \dt\ is calculated from the measured separation \dz\ between
the decay vertices of $B_{rec}$ and $B_{tag}$ along the collision axis ($z$).
The vertex of $B_{rec}$ is reconstructed from the lepton tracks that come from the $J/\psi$;
the vertex of $B_{tag}$ is constructed from the remaining tracks in the event that do
not belong to $B_{rec}$, with constraints from the beam spot location
and the $B_{rec}$ momentum.  We accept events with $|\dt|<20 \ps$ whose
uncertainty are less than $2.5 \ps$.  

\par After all of the selection criteria mentioned above have been applied, the average
number of candidates per event is approximately 1.1, indicating some events still have multiple candidates.
In these events, we randomly choose one candidate to be used in the fit.
%This selection is unbiased and correctly identifies the true signal candidate $91.7 \%$ of the time.
This selection is unbiased. Overall, the true signal candidate is correctly identified $91.7 \%$ of the time.
After this step, the signal efficiency is $22.0 \%$ and a total of 1318 on-peak events are selected.

\par In addition to signal and continuum background events, there are also \bb-associated backgrounds present 
in the data.  We divide the \B\ backgrounds into the
following types: (i) \btojpsiks, where \kstopizpiz\, (ii) inclusive neutral \B\ meson decays, and (iii) inclusive charged
\B\ meson decays. When normalized to the integrated luminosity, Monte Carlo (MC) studies predict $153\pm 9$, 
$68 \pm 14$ and $314 \pm 63$ events of these background types, respectively. The inclusive  neutral \B\ meson decays exclude
signal and \btojpsiks\ events. The inclusive \B\ decay backgrounds are dominated by contributions from
$\B \to J/\psi X$ (inclusive charmonium final states). In particular the inclusive charged
\B\ meson decay backgrounds are dominated by $\B^{\pm} \to J/\psi \rho^{+}$ decays. The \btojpsiks\ background was 
studied separately since its \CP asymmetries are precisely measured.

\par The signal yield, \S\ and \C\ are simultaneously extracted from an unbinned maximum-likelihood (ML)
fit to the $\B$ candidate sample, where the discriminating variables used in the fit are 
\mes, \DeltaE, \fish\ and \dt. The continuum yield for the seven mutually-exclusive tagging 
categories, is also allowed to vary in the ML fit.

The probability density function (PDF) for signal \mes\ distribution takes the form of a
Gaussian with a low side exponential tail~\cite{ref:crystalball}.
We parameterize the \mes\ distribution for continuum 
and neutral inclusive \B\ background with an Argus phase space distribution~\cite{ref:argus}.
As there are significant correlations between \mes\ and \DeltaE for the 
charged inclusive \B\ and the \btojpsiks\ backgrounds, we parameterize these variables with
two-dimensional non-parametric PDFs. 
%We also use two-dimensional non-parametric PDFs when describing the \mes-\DeltaE distribution for \btojpsiks.
The \DeltaE distribution for signal events is modeled by a Gaussian with an
exponential tail on the negative side to account for energy leakage in the EMC, 
plus a polynomial contribution. The \DeltaE distributions for the 
continuum and the neutral inclusive \B\ background are described by second and third-order
polynomials respectively. The \fish\ distributions for the signal and the backgrounds are described by 
bifurcated Gaussians with different widths above and below the peak value.

The signal decay rate distribution of Equation~\ref{equation-ff} is modified to account for dilution coming from 
incorrectly assigning the flavor of $B_{tag}$ and is convolved with a triple 
Gaussian resolution function, whose core width is about 1.1 \ps~\cite{ref:bigprd}.
The decay rate distribution for \B\ backgrounds is similar to that for signal.
%To account for their mis-reconstruction, the inclusive \B\ backgrounds are assigned 
%an effective lifetime instead of their respective measured \B\ lifetimes.
The inclusive \B\ backgrounds are assigned an effective lifetime instead of their
respective \B\ lifetimes to account for their mis-reconstruction. This effective
lifetime is determined from MC simulated data. 
%When evaluating systematic uncertainties, we allow for \CP\ violation 
%in the inclusive \B background. This is described later in the text. 
The potential \CP\ asymmetry of the inclusive \B\ background is evaluated by allowing
the parameters of \S\ and \C\ for this background to vary.
The decay rate distribution for $\btojpsiks$ is the same as that for signal 
and reflects the known level of \CP\ violation in that decay.  The continuum 
background is modeled with a prompt lifetime component convolved with a triple 
Gaussian resolution function. The core Gaussian parameters and fractions 
are allowed to vary in the ML fit. The other two Gaussians have means fixed 
to zero, and widths of 0.85 \ps and 8.0 \ps, respectively.

The results from the ML fit are 109 $\pm$ 12 \stat signal events, with \S\ = $-$0.68 $\pm$ 0.30 \stat
and \C = $-$0.21 $\pm$ 0.26 \stat. 
The fit yields the following numbers of
continuum events: N$_{Lepton}$ = 17 $\pm$ 5,
N$_{KaonI}$ = 38 $\pm$ 8, N$_{KaonII}$ = 101 $\pm$ 12, 
N$_{KaonPion}$ = 102 $\pm$ 12, N$_{Pion}$ = 115 $\pm$ 12, 
N$_{Other}$ = 94 $\pm$ 11 and N$_{NoTag}$ = 227 $\pm$ 17.
%--------------------------------------------------------------------
% Figure : Projection plots for mes, deltaE and fisher from Run 1-4
%--------------------------------------------------------------------
\begin{figure}[ht]
\begin{center}
\includegraphics[height=5.0cm]{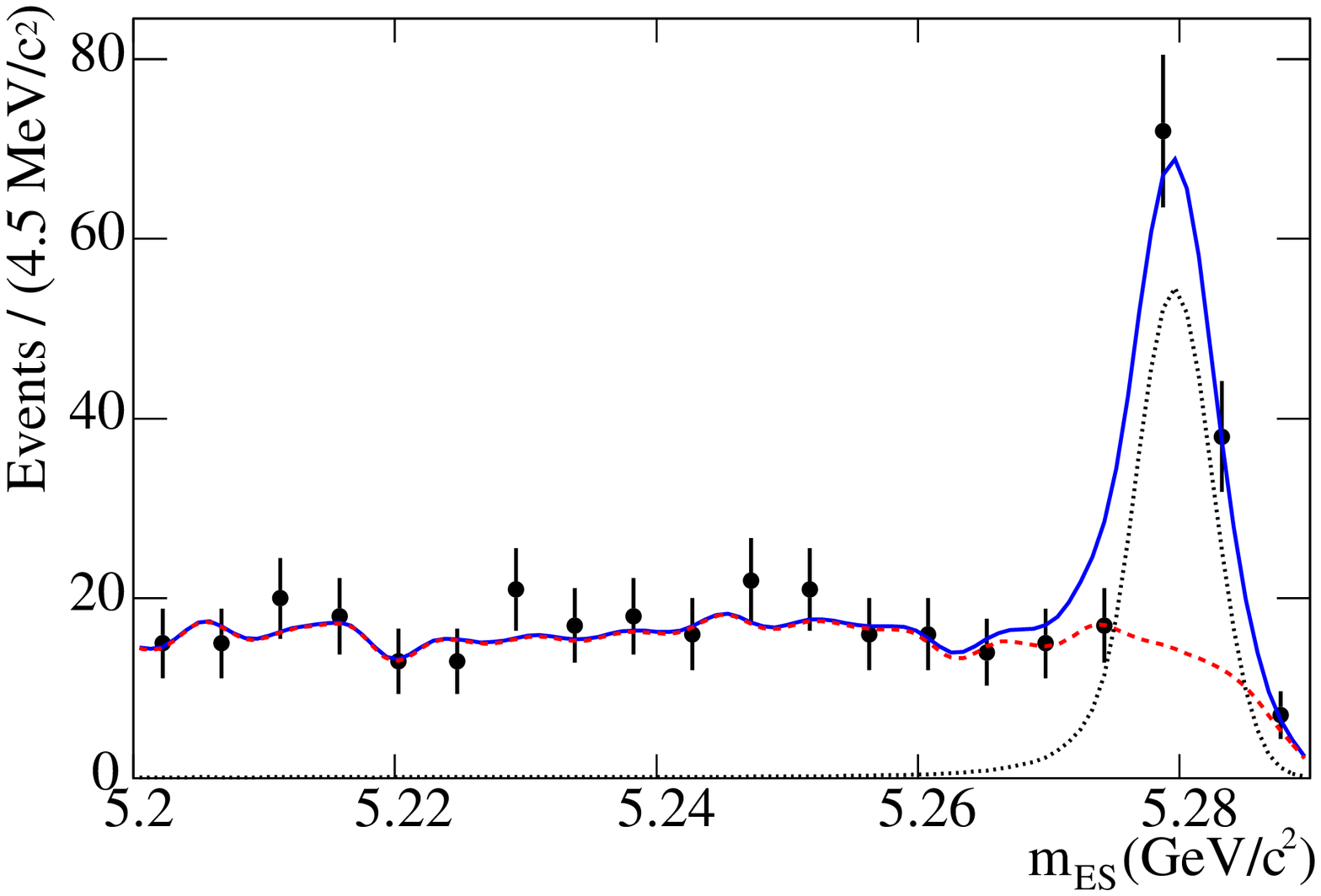}
\includegraphics[height=5.0cm]{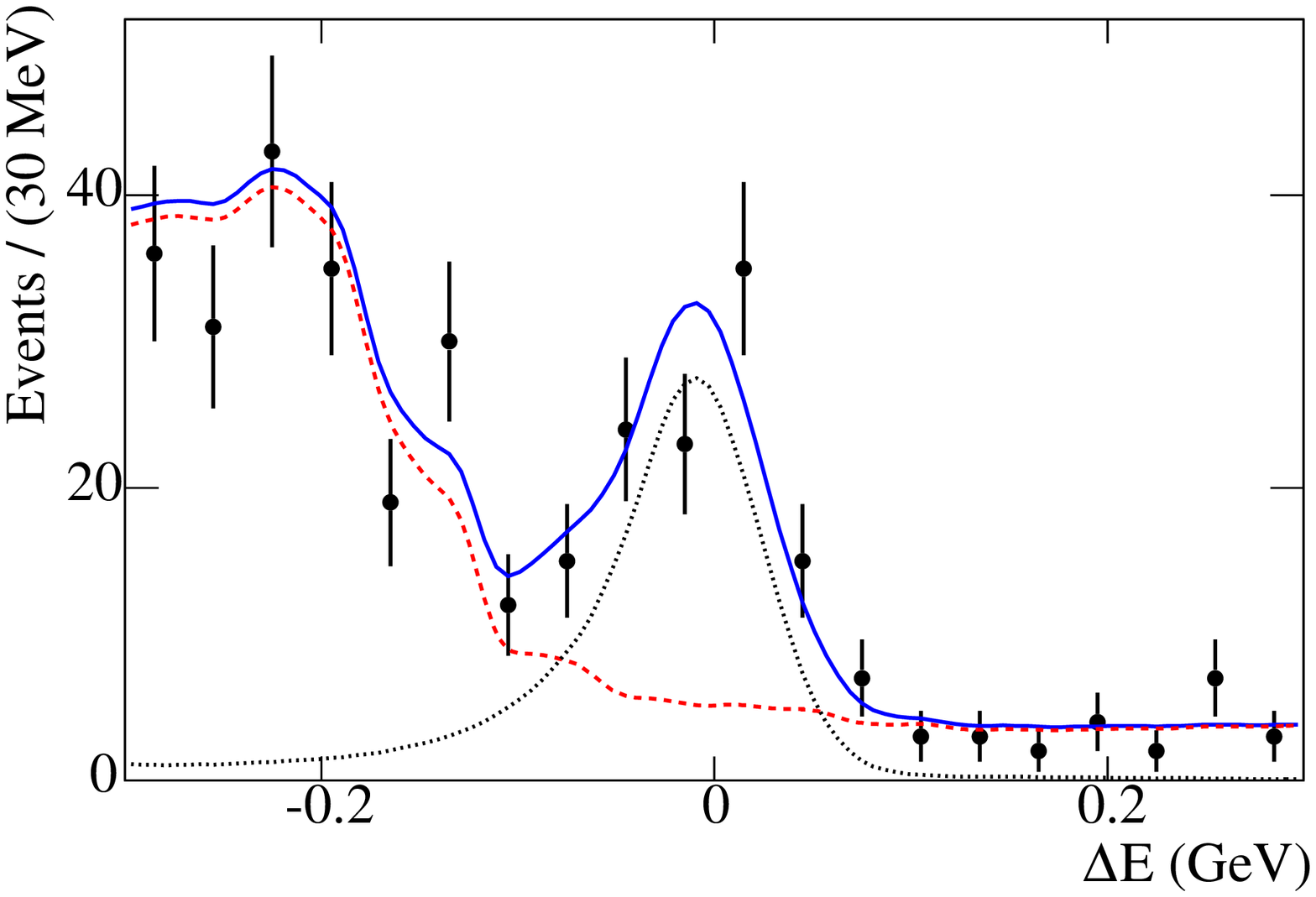}
\includegraphics[height=5.0cm]{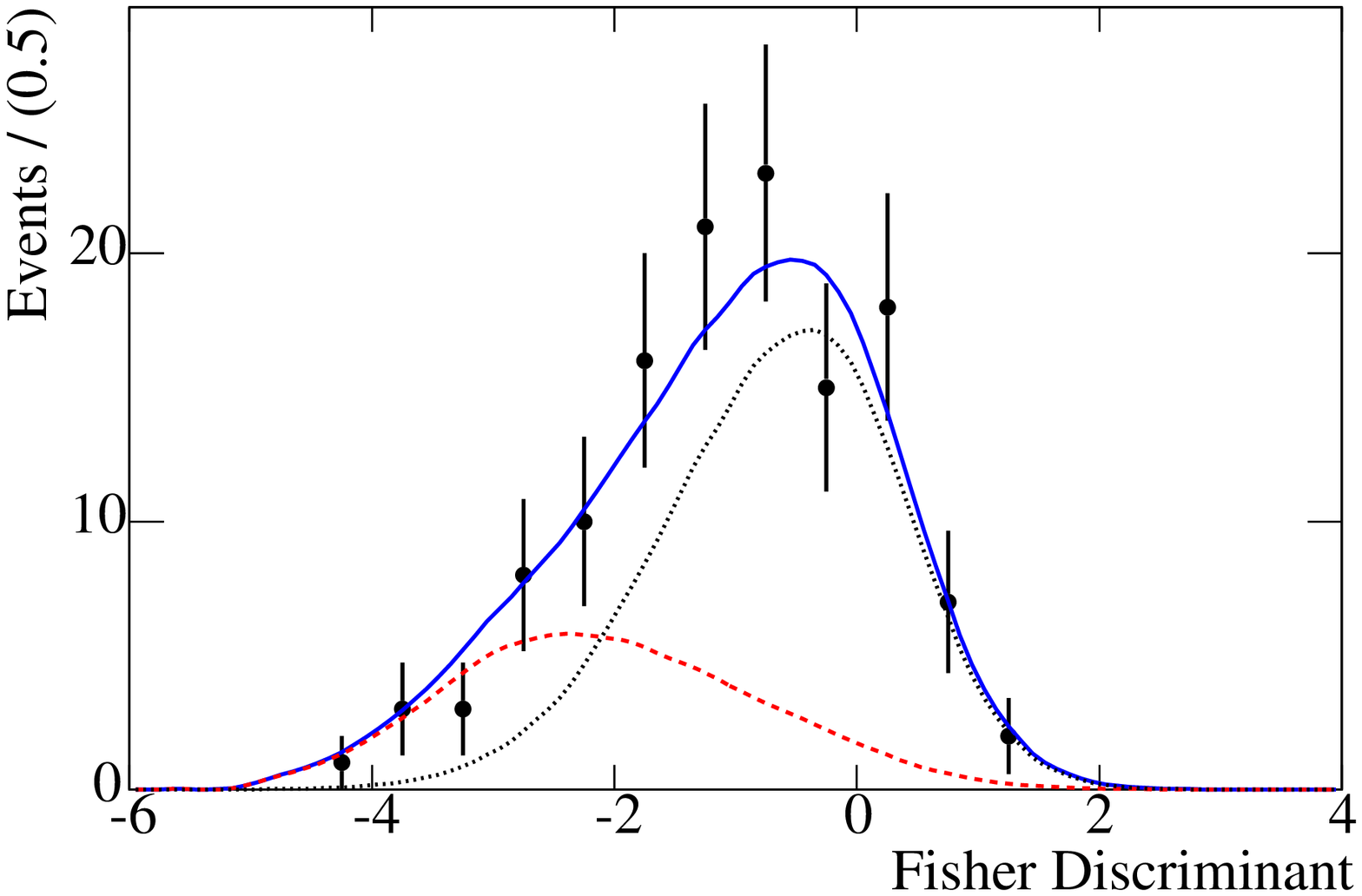}
\caption{Signal enhanced distributions of \mes\ (top), \DeltaE (center) and
 \fish\ (bottom) for the data (points). The solid line represents the 
total likelihood, the dashed line is the sum of the backgrounds and the 
dotted line is the signal. The undulations in the background model are the 
result of limited MC statistics available for defining the two-dimensional 
non-parametric PDFs.}
\label{fig:projection}
\end{center}
\end{figure}
%------------------------------
%-------------------------------------------------
% Figure : Asymmetry plots
%-------------------------------------------------
\begin{figure}[ht]
\begin{center}
\includegraphics[height=6.5cm]{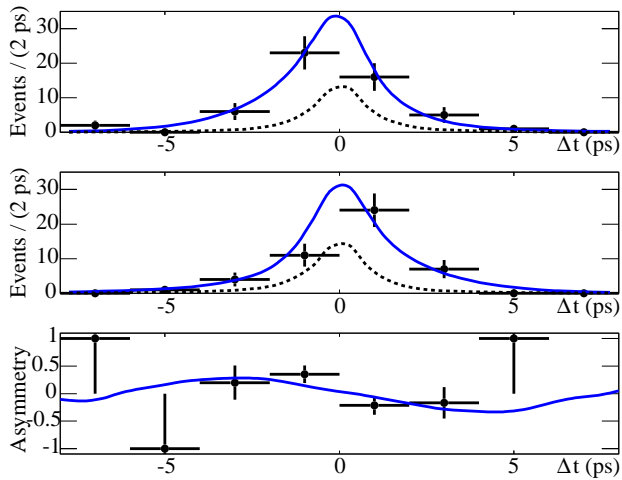}
%\caption{The \deltat\ distribution for a sample of signal enhanced events for $\Bz$ (top) and $\Bzb$ (middle) tagged events. 
\caption{The \deltat\ distribution for a sample of signal enhanced events tagged as $\Bz$ (top) and $\Bzb$ (middle). 
The dotted lines are the sum of backgrounds and the solid lines are the sum of signal and backgrounds.  
The time-dependent \CP asymmetry (see text) is also shown (bottom), where the curve is the measured asymmetry.}
\label{fig:asym}
\end{center}
\end{figure}
%------------------------------
Figure~\ref{fig:projection} shows the distributions of \mes, \DeltaE, and
\fish\ for the data.  In these plots the signal has been enhanced by 
selecting $|\DeltaE| < 0.1 \gev$ for the \mes\ plot, 
$\mes > 5.275\gevcc$ for the \DeltaE plot and by applying
 both of these criteria for the \fish\ plot.  
After applying these requirements to the signal (background) samples that
 are used in the fit, they are reduced to a relative size of 83.1\% (24.3\%), 
85.0\% (21.1\%) and 73.1\% (2.8\%) for the \mes, \DeltaE, and \fish\ distributions respectively.

Figure~\ref{fig:asym} shows the \deltat\ distribution for signal \Bz and \Bzb
tagged events. The signal has been enhanced using the same \mes\ and \DeltaE\ cuts
as for Figure~\ref{fig:projection}. The time-dependent decay rate asymmetry 
$[  N (\deltat) - \overline{N}(\deltat) ] / [  N (\deltat) + \overline{N}(\deltat) ]$
is also shown, where $N$ $(\overline{N})$ is the decay rate for \Bz(\Bzb) tagged events
and the decay rate takes the form of Equation~\ref{equation-ff}.

Table~\ref{table:systematicerrors} summarizes the systematic uncertainties on 
the signal yield, \S\ and \C. 
%Each entry in the table indicates one systematic effect and these are added in quadrature to give the totals presented. 
These include the uncertainty due to the PDF parameterization (including the resolution function), 
evaluated by varying the signal and the background PDF parameters within uncertainties of their nominal values.
%by fixing both the signal and the background PDF parameters to their nominal values
%and varying them within uncertainties;
the effect of SVT mis-alignment; the uncertainties associated with the Lorentz boost, the
z-scale of the tracking system, and the event-by-event beam spot position. 

%------------------------------------------------------------------------------------------------------------------
% Table : Systematic errors
%------------------------------------------------------------------------------------------------------------------
\begin{table}[h]
\caption{Contributions to the systematic errors on the signal yield, \S\ and \C, where the
signal yield errors are given in numbers of events. The total systematic
uncertainty is the quadratic sum of the individual contributions listed. Additional systematic 
uncertainties that are applied only to the branching fraction are discussed in the text.~\label{table:systematicerrors}}
\begin{center}
\begin{tabular}{|c|c|c|c|}\hline
Contribution                    & Signal yield       & \S\                  & \C\                 \\\hline\hline
                                &                    &                      &                     \\
PDF parameterization            & $^{+3.21}_{-2.88}$ & $\pm${0.013}         & $\pm${0.012}        \\
SVT mis-alignment               & $-$                & $\pm${0.002}         & $\pm${0.002}        \\
Boost and z-scale               & $^{+0.08}_{-0.16}$ & $\pm${0.004}         & $\pm${0.001}        \\
Beam spot position              & $-$                & $\pm${0.007}         & $\pm${0.002}        \\
Fit bias                        & $\pm$3.00          & $\pm${0.026}         & $\pm${0.016}         \\
Inclusive $\B$ background yields  & $\pm$3.52        & $\pm${0.003}         & $\pm${0.020}         \\ 
$\mes$-$\DeltaE$ correlations   & $\pm$2.92          & $\pm${0.020}         & $\pm${0.002}        \\
\CP\ content of \B\ background  & $^{+0.13}_{-0.11}$ & $\pm${0.012}         & $\pm${0.049}        \\
\CP\ background lifetime        & $\pm$0.67          & $\pm${0.010}         & $\pm${0.010}         \\
Tagging efficiency asymmetry    & $\pm$0.02          & $\pm${0.000}         & $\pm${0.020}         \\
Tag-side interference           & $-$                & $\pm${0.004}         & $\pm${0.014}        \\
Fisher data/MC comparison       & $\pm${0.70}        & $\pm${0.004}         & $\pm${0.004}        \\\hline      
                                &                    &                      &                     \\
Total                           & $^{+6.42}_{-6.26}$ & $\pm${0.040}          & $\pm${0.063}         \\
                                &                    &                      &                     \\\hline
\end{tabular}
\end{center}
\end{table}
%------------------------------------------------------------------------------------------------------------------
\par The uncertainty coming from the fit bias is estimated by performing ensembles of mock 
experiments using signal MC which is generated using the {\tt GEANT4}-based~\cite{ref:geant} \babar\ 
MC simulation, embedded into MC samples of background generated from the 
likelihood.  The deviation from input values is added in quadrature to the error on the 
deviation in order to obtain a conservative fit bias uncertainty. Most, but not all of the inclusive 
charmonium final states which dominate the inclusive \B\ background, are precisely known 
from previous measurements. Their yields are then fixed in the fit. 

As a crosscheck, the yields for inclusive \B\ backgrounds that are not well known are allowed 
to vary. The deviation from the nominal result is taken as a systematic uncertainty. We 
include an additional systematic uncertainty to account for neglecting the small correlation 
between $\mes$ and $\DeltaE$ in signal and neutral inclusive \B\ background events.

\par In order to evaluate the uncertainty coming from \CP\ violation in the \B\ background,
we have allowed the \S\ and \C\ parameters to vary in a fit for the
neutral inclusive \B\ background, and have separately allowed the \C\ parameter
to vary in a fit for the charged inclusive \B\ background. The deviations of the 
fitted values of the signal \S\ and \C\ from the nominal fit results are assigned as systematic 
errors. The uncertainty from \CP\ violation in \btojpsiks\ is determined by varying \S\ and 
\C\ within current experimental limits~\cite{ref:babar2004}. 

\par The inclusive \B\ background uses an effective lifetime in the nominal fit and
we replace this with the world-average \B\ lifetime~\cite{ref:pdg2004} 
to evaluate the systematic error due to the \CP\ background lifetime.
There is also a small asymmetry in the tagging efficiency between $\Bz$ and $\Bzb$ tagged events, for
which a systematic uncertainty is evaluated. We study the possible interference between the 
suppressed $\bar b\to \bar u c \bar d$ amplitude with the favored $b\to c \bar u d$ amplitude 
for some tag-side $B$ decays~\cite{ref:dcsd}. The difference in the distribution of \fish\
between data and MC is evaluated with a large sample of $\btodstarrho$ decays.
There are additional systematic uncertainties that contribute only to the branching fraction.  These come from
uncertainties for charged particle identification (5.2\%), \piz\ meson reconstruction efficiency (3\%), the $\jpsitoll$
branching fractions (2.4\%), the tracking efficiency (1.2\%) and the number of \B\ meson
pairs (1.1\%). The systematic error contribution from MC statistics is negligible.  
The 109 $\pm$ 12 signal events correspond to a branching fraction of
\vspace{-0.8cm}
\begin{center}
$$
{\cal{B}}(\btojpsipi) = (1.94 \pm 0.22 \stat \pm 0.17 \syst)\times 10^{-5}.
$$
\end{center}\vspace{-0.2cm}
We determine the \CP asymmetry parameters to be
\vspace{-0.9cm}
\begin{center}
\begin{eqnarray}
\C\ = -0.21 \pm 0.26 \stat \pm 0.06\syst, \nonumber \\
\S\ = -0.68 \pm 0.30 \stat \pm 0.04\syst, \nonumber
\end{eqnarray}
\end{center}\vspace{-0.2cm}
where the correlation between \S\ and \C\ is 8.3\%. The value of \S\ is consistent with
SM expectations for a tree-dominated ${b \rightarrow c\mskip 2mu \overline c \mskip 2mu d}$ 
transition of \S\ = $-\sintwob$ and \C\ = 0. All results presented here are consistent
with previous measurements from the \B\ Factories~\cite{ref:babarbf,ref:babarcp,ref:bellebf,ref:bellecp}.

% Input the pubboard acknowledgements file
% \input pubboard/acknow_PRL.tex
%\input pubboard/acknowledgements.tex
We are grateful for the 
extraordinary contributions of our \pep2\ colleagues in
achieving the excellent luminosity and machine conditions
that have made this work possible.
The success of this project also relies critically on the 
expertise and dedication of the computing organizations that 
support \babar.
The collaborating institutions wish to thank 
SLAC for its support and the kind hospitality extended to them. 
This work is supported by the
US Department of Energy
and National Science Foundation, the
Natural Sciences and Engineering Research Council (Canada),
Institute of High Energy Physics (China), the
Commissariat \`a l'Energie Atomique and
Institut National de Physique Nucl\'eaire et de Physique des Particules
(France), the
Bundesministerium f\"ur Bildung und Forschung and
Deutsche Forschungsgemeinschaft
(Germany), the
Istituto Nazionale di Fisica Nucleare (Italy),
the Foundation for Fundamental Research on Matter (The Netherlands),
the Research Council of Norway, the
Ministry of Science and Technology of the Russian Federation, and the
Particle Physics and Astronomy Research Council (United Kingdom). 
Individuals have received support from 
CONACyT (Mexico), the Marie-Curie Intra European Fellowship program (European Union),
the A. P. Sloan Foundation, 
the Research Corporation,
and the Alexander von Humboldt Foundation.

%-----------------------------
% Bibliography
%-----------------------------

%-------------------------------------------


\begin{thebibliography}{99}

% [1] BaBar establishes CP using sin2beta
\bibitem{babar-stwob-prl}
\babar\ Collaboration, B.\ Aubert {\em et al.},
\jprl{89}, 201802 (2002).

% [2] Belle establishes CP using sin2beta
\bibitem{belle-stwob-prl}
BELLE Collaboration, K.\ Abe {\em et al.},
\jprd{66}, 071102 (2002).

% [3] CKM matrix
\bibitem{ref:CKM}
N.~Cabibbo, Phys.~Rev.~Lett.~{\bf 10}, 531 (1963);\\
M.~Kobayashi and T.~Maskawa, Prog.\ Th.\ Phys.\ {\bf 49}, 652 (1973).

% [4] Direct measurement of sin(2\beta)
\bibitem{BCP}
A.B.~Carter and A.I.~Sanda, \pr {\bf D23}, 1567 (1981);\\
I.I.~Bigi and A.I.~Sanda, \np {\bf B193}, 85 (1981).

% [5] Is time-dependent asymmetry the same for b-> c cbar d as for b->c cbar s ?
\bibitem{grossman}
Y.~Grossman and M.~Worah,
\plb{395}, 241 (1997).

% [6] The Effect of Penguins in the \btojpsiks CP Asymmetry
\bibitem{ciuchini}
M.~Ciuchini, M.~Pierini and L.~Silvestrini,
\jprl{95}, 221804 (2005).

% [7] BaBar J/psi pi0 branching fraction
\bibitem{ref:babarbf}
\babar\ Collaboration, B.\ Aubert {\em et al.},
\jprd{65}, 032001 (2002).

% [8] BaBar J/psi pi0 CP
\bibitem{ref:babarcp}
\babar\ Collaboration, B.\ Aubert {\em et al.},
\jprl{91}, 061802 (2003).

% [9] Belle J/psi pi0 branching fraction (29.4 fb-1)
\bibitem{ref:bellebf}
BELLE Collaboration, K.\ Abe {\em et al.},
\jprd{67}, 032003 (2002).

% [10] Belle J/psi pi0 CP
\bibitem{ref:bellecp}
BELLE Collaboration, K.\ Abe {\em et al.},
\jprl{93}, 261801 (2004).

% [11] The \babar\ detector and dataset : NIM detector performance paper
\bibitem{ref:babar}
\babar\ Collaboration, B.\ Aubert {\em et al.},
\nima{479}, 1 (2002).

% [12] 2004 PDG
\bibitem{ref:pdg2004}
Particle Data Group, 
S.~Eidelman {\em et al.},
\plb{592}, 1 (2004).

% [13] Fisher discriminant
\bibitem{ref:fisher}
R.~A.~Fisher, Annals of Eugenics 7, 179 (1936).

% [14] Summer 2004 sin2\beta PRL
\bibitem{ref:babar2004}
\babar\ Collaboration, B.\ Aubert {\em et al.},
\jprl{94}, 161803 (2005).

% [15] Crystal Ball function
\bibitem{ref:crystalball}
Crystal Ball Collaboration, D.~Antreasyan {\em et al.}, Crystal Ball Note 321 (1983).

% [16] Argus function
\bibitem{ref:argus}
ARGUS Collaboration, H.~Albrecht {\em et al.},
\plb{241}, 278 (1990).

% [17] BaBar big sin2beta PRD
\bibitem{ref:bigprd}
\babar\ Collaboration, B.\ Aubert {\em et al.},
\jprd{66}, 032003 (2002).

% [18] GEANT-4
\bibitem{ref:geant}
{\tt GEANT4} Collaboration, S.~Agostinelli {\em et al.},
\nima{506}, 250 (2003).

% [19] DCSD
\bibitem{ref:dcsd}
O.~Long, M.~Baak, R.~N.~Cahn, D.~Kirkby, 
\jprd{68}, 034010 (2003).

\end{thebibliography}
\end{document}